	\documentclass[12pt, a4paper]{article}
	
 	\usepackage{latexsym}
	\usepackage{amsmath}	
	\usepackage{epsfig}	
	\usepackage{times}	
	\usepackage{amssymb}
	
 	\newtheorem{thr}{Theorem}[section]
 	\newtheorem{lem}{Lemma}[section]
 	\newtheorem{prop}{Proposition}[section]
 	
	\newtheorem{cor}{Corollary}[section]
	\newtheorem{que}{Question}

 	\newcommand{\F}{\mathbb{F}}
 	
 	\newcommand{\RS}{\textnormal{RS}}
 	\newcommand{\SSRS}{\textnormal{SSRS}}
 	
\begin{document}
 	
 	\centerline{\Large{\bf Some New Results on Equivalency of Collusion-Secure}}
 	\centerline{}
 	\centerline{\Large{\bf Properties for Reed-Solomon Codes}}
 	\centerline{}
 	\centerline{}
 	\centerline{Myong-Son Sin and Ryul Kim}
 	\centerline{}
 	{\small \centerline{{\it faculty of Mathematics, \textbf{Kim Il Sung} University,}}}
	{\small \centerline{{\it Pyongyang, D.P.R Korea}}}

	\begin{abstract}
	A. Silverberg (IEEE Trans. Inform. Theory 49, 2003)
	proposed a question on the equivalence of identifiable parent
	property and traceability property for Reed-Solomon code family.
	Earlier studies on Silverberg's problem motivate us to think of
	the stronger version of the question on equivalence of separation
	and traceability properties. Both, however, still remain open.
	In this article, we integrate all the previous works on this problem with an algebraic way,
	and present some new results. It is notable that
	the concept of subspace subcode of Reed-Solomon code, which was introduced in
	error-correcting code theory, provides an interesting prospect for our topic.
	\end{abstract}
	\section{Introduction}
	The growth of Internet raised the problem of illegal redistribution as a
	major concern in digital content industry, because copying such material
	is easy and no information is lost in the process. To protect digital copies,
	however, is a complicated	task. Methods like cryptography do not resolve this
	problem, since the information must be decrypted at one point to be able to
	use it. The goal of digital fingerprinting is to discourage people from
	illegally redistributing their legally purchased copy. In this scenario, the
	distributor embeds into the digital content, using a watermark algorithm, a
	unique piece of information({\it{fingerprint}}) for each user. If an illegal copy is
	found, the distributor can extract the fingerprint from it to identity the
	dishonest user({\it{pirate}}). Because the pirate may try to damage the fingerprint
	before redistribution, the watermarking algorithm must ensure robustness to the distributor.
	
	Nevertheless, the most dangerous attack against digital fingerprinting is the
	collusion attack introduced in \cite{bon}. The contents delivered to different
	users are, since their
	fingerprints differ, essentially different. Two or more pirates may compare
	their copies and reveal the locations of part of fingerprint. With deleting or
	modifying those locations, pirates can generate a new copy of content in order
	not to be traced. This collusion attack could not only violate
	pirate-identifying but frame an innocent user in some cases. We are interested
	in designing a set of fingerprints({\it{fingerprinting code}}) with which the
	distributor can always identity at least one colluder from a forged fingerprint
	with a small error probability. In particular separating code, IPP code and TA code are
	most important fingerprinting codes with different
	collusion-secure properties for generic digital data.
	
	We will denote the $i$th component of any tuple $x$ by $x_i$ and the Hamming
	distance between two tuples $x, y$ by $d(x,y)$. Let $n,w,w_1$ and $w_2$ be positive
	integers such that $n,w,w_1 \geq 2$ and $w_1 \geq w_2$. Suppose $C$ is a code of length
	$n$ over $\F_q$.
	\begin{itemize}
	\item We define \textit{descendant set} of an arbitrary nonempty subset $U$ of $C$ by
	\begin{equation*}
	\textnormal{desc}U:=\{x \in \F_q^n \mid \forall i, \exists y \in U : x_i=y_i \}
	\end{equation*}
	\item $C$ is a $(w_1,w_2)$-\textit{separating code} provided that, if $U_1,U_2$ are
	disjoint subsets of $C$ such that $1 \leq |U_1| \leq w_1$ and $1 \leq |U_2| \leq w_2$,
	then their descendant sets are also disjoint.
	\item $C$ is called a $w$-\textit{identifiable parent property code}(IPP code)
	provided that for all $x \in \F_q^n$, the set $\textrm{IPP}_w(x):=\{U \subset C \mid
	x \in \textrm{desc}U, 1 \leq |U| \leq w \}$ is empty or
	$\bigcap \limits_{U \subset \textrm{IPP}_w(x)}U \neq \phi$.
	\item $C$ is called a $w$-\textit{traceability code}(TA code)	provided that if
	$U \subset C, 1 \leq |U| \leq w$ and $x \in \textrm{desc}U$, there exist at
	least one codeword $y \in U$ such that $d(x,y)<d(x,z)$ for all $z \in C \backslash U$.
	\end{itemize}	
	The code classes defined above are known to satisfy the following relationships.
	\begin{prop}
	\textnormal{(see \cite{ssw01})} Let $d$ be the minimum distance of a code $C$ of
	length $n$. Then for $C$, \\
	\centerline{
	$d>n(1-1/w^2) \Rightarrow w$-TA $\Rightarrow w$-IPP $\Rightarrow (w,w)$-separating}
	\end{prop}
	\begin{prop}
	\textnormal{(see \cite{c})} Let $d$ be the minimum distance of a code $C$ of
	length $n$. If $d>n(1-1/(w_1w_2))$, then $C$ is a $(w_1,w_2)$-separating code.
	\end{prop}
	Let $1 \leq k \leq q-1$ be an integer. The \textit{Reed-Solomon code} $\RS_k(q)$
	of dimension $k$ over $\F_q$ is defined by $\RS_k(q):=\{ev(f) \mid f \in \F_q[x],
	\textrm{deg}f<k\}$,	where $ev:f \in F_q[x] \mapsto (f(\alpha^0),f(\alpha^1),
	\cdots,f(\alpha^{q-2}))	\in \F_q^{q-1}$	and $\alpha$ is a primitive element in $\F_q$.
	It is well known that $\RS_k(q)$ is a $[q-1,k,q-k]$-linear code.	
	\par
	Reed-Solomon code is one of the most famous error-correcting codes and
	it also has an application in digital fingerprinting. A. Silverberg, et al.
	\cite{ssw03} dealt with applying list decoding method to tracing algorithms
	of fingerprinting codes. In their work, the collusion-secure properties of
	Reed-Solomon codes and other algebraic geometry codes were studied, and
	the following question was left as an open problem.
	\begin{que}
	Is it the case that $d>n-n/w^2$ for all $w$-IPP Reed-Solomon codes of length $n$
	and minimum distance $d$?
	\end{que}
	Thus,	Silverberg's question is a problem of the equivalence of IPP and traceability
	for Reed-Solomon code family. \\
	\indent The problem was studied in \cite{fcsd} and \cite{mfs}. In \cite{fcsd}, they
	restated the separation property of Reed-Solomon codes algebraically, as a system of
	equations, to get the following result.
	\begin{thr} \textnormal{(see \cite{fcsd})}
	Suppose $k-1$ divides $q-1$. If $\RS_k(q)$ is a $(w_1,w_2)$-separating code,
	then $d>n-n/(w_1w_2)$ where $d$ is minimum distance.
	\end{thr}
	In \cite{mfs}, they presented the similar result as follows by establishing an
	additive homomorphism over finite field.
	\begin{thr} \textnormal{(see \cite{mfs})}
	Suppose $w^2>q$ or $w$ divides $q$. If $\RS_k(q)$ is a $(w,w)$-separating code,
	then $d>n-n/w^2$ where $d$ is minimum distance.
	\end{thr}
	As you can see, the previous works claimed the stronger fact than Silverberg's original
	problem in certain cases. In this context, we naturally raise the following question,
	which turns out to be the main topic of this article.
	\begin{que}
	Is it the case that $d>n-n/(w_1w_2)$ for all $(w_1,w_2)$-separating Reed-Solomon
	codes of length $n$	and minimum distance $d$?
	\end{que}
	The rest of the paper is organized as follows: In Section 2, we will present a
	sufficient condition for non-separation of linear codes, and prove that the previous
	works can be derived from that condition. Some more parameter setups providing
	positive answer about Question 2 will be obtained in Section 3. In Section 4, the
	application of subspace subcodes of Reed-Solomon codes will be unveiled. We conclude
	the paper in Section 5 after presenting experimental results to show the extension
	of our work.
	\par Throughout the remaining, $\F_q$ is Galois field with order $q=p^m$ and
	characteristic $p$. Let $r_i=[\textnormal{log}_pw_i], i=1,2$. For any polynomial
	$f$ over $\F_q$, let Im$f=f(\F_q)$. For an arbitrary word $x \in \F_q^n$, Im$x$
	is the set of all its components, i.e. Im$x=\{x_i \mid 1 \leq i \leq n\}$.
	For given two sets $E,F \subset \F_q$, we define $EF:=\{ab \mid a \in E,
	b \in F\}$ and $E+F:=\{a+b \mid a \in E, b \in F\}$. We will denote the set of
	all polynomials over $\F_q$ of degree less than $k$ by $P_k$. $n,w,w_1$ and
	$w_2$ are positive integers satisfying $n,w,w_1 \geq 2$ and $w_1 \geq w_2$.
	\section{Restatement of the Previous Works}
	In this section we propose a sufficient condition for non-separation of linear
	codes, which will integrate the former results in \cite{fcsd} and \cite{mfs}.
	The idea was motivated by \cite{mfs}, where an additive homomorphism was
	established such that its image set has a special property. Before presenting
	the major result, we will formally define such "special property" of a set. \par
	\indent Let $U$ be a subset of $\F_q$. $U$ is called \textit{additively}
	(\textit{multiplicatively}) $(w_1,w_2)$- \textit{separable} and written by
	$U=(E,F)_{w_1,w_2}$ provided that
	there exist two subsets $E,F \subset U$ with $1 \leq |E| \leq w_1$ and
	$1 \leq |F| \leq w_2$ such that $U \subset E+F$ ($U \subset EF$). \par
	\indent The following theorem is the main result of this section. Note that
	it is not just for Reed-Solomon codes, but for linear codes.
	\begin{thr}
	Let $C$ be $[n,k]_q$-linear code containing \textnormal{\textbf{1}}$=(1,1,\cdots,1)$.
	Suppose there exists a codeword $c=(c_1,c_2,\cdots,c_n) \in C$ such that
	$|\textnormal{Im}c| \geq 2$ and \textnormal{Im}$c$ is $(w_1,w_2)$-separable additively
	or multiplicatively. Then, $C$ is not $(w_1,w_2)$-separating.
	\end{thr}
	\textit{Proof}. We will only prove when Im$c$ is additively $(w_1,w_2)$-separable,
	since the other case can be proven in similar way. Let Im$c=(E,F)_{w_1,w_2}$.
	Define $U:=\{\beta \cdot \textbf{1} \mid \beta \in E\}$ and
	$V:=\{c-\gamma \cdot \textbf{1} \mid \gamma \in F\}$. Then $U,V \subset C$ since
	$c,\textbf{1} \in C$ and $C$ is a linear code. Further, $U$ and $V$ are disjoint because
	$|\textnormal{Im}c| \geq 2$.
	For all $i \in \overline{1,n}$, there exist $\beta_i \in E$ and $\gamma_i \in F$ such that
	$c_i=\beta_i+\gamma_i$. If we set $x:=(\beta_1,\beta_2,\cdots,\beta_n)$, it is clear
	that $x \in \textrm{desc}U \cap \textrm{desc}V$ which implies non-separation. $\boxdot$ \par
	\bigskip
	\noindent The following corollary	is the Reed-Solomon code version of Theorem 2.1.
	\begin{cor}
	Let $1 \leq k \leq q-1$ be an integer. If there exists a non-constant polynomial $f$
	in $P_k$ such that \textnormal{Im}$f$ is $(w_1,w_2)$-separable additively or
	multiplicatively, then the code $\RS_k(q)$ is not $(w_1,w_2)$-separating.
	\end{cor}
	\noindent By definition, $\RS_k(q) \subset \RS_{k+1}(q)$, therefore the code $\RS_{k+1}(q)$
	is not $(w_1,w_2)$-separating if $\RS_k(q)$ is not $(w_1,w_2)$-separating. Meanwhile,
	the inequality $d>n-n/(w_1w_2)$ is equivalent with $k-1<(q-1)/(w_1w_2)$ for
	Reed-Solomon codes. Thus, it sufices to consider the case
	$k=\lceil (q-1)/(w_1w_2) \rceil + 1$ when we study Question 2.
	In other words, if $\RS_k(q)$ is not $(w_1,w_2)$-separating where
	$k=\lceil (q-1)/(w_1w_2) \rceil + 1$ for given $q,w_1,w_2$, Question 2 has the positive
	answer. (see \cite{mfs}) \par
	\indent In this context, we will reprove the previous results done on Silverberg's
	open problem more simply using Corollary 2.1. \par \bigskip
	\indent \textit{Proof of Theorem 1.1 :} Suppose $d \leq n(1-1/(w_1w_2))$, i.e.
	$k-1 \geq (q-1)/(w_1w_2)$. Set $f(x):=x^{k-1}$. Since $k-1 \mid q-1$,
	the polynomial $f$ is a multiplicative homomorphism mapping $\F_q^{*}$ to $\F_q^{*}$.
	So Im$f$ is a multiplicative subgroup of $\F_q^*$ with order
	$|\textrm{Im}f|=|\F_q^*|/|\textrm{Ker}f|=(q-1)/(k-1) \leq w_1w_2$. For $\F_q^*$ is cyclic,
	Im$f$ is also cyclic, thus, it has a generator $\gamma$. Set
	$E:=\{\gamma^{iw_2} \mid 0 \leq i \leq w_1-1 \}$ and $F:=\{\gamma^j \mid 0 \leq j \leq w_2-1 \}$.
	Then it is easy to check that Im$f=(E,F)_{w_1,w_2}$, which implies non-separation by
	Corollary 2.1.
	Therefore, if $\RS_k(q)$ is $(w_1,w_2)$-separating, then $k-1<(q-1)/(w_1w_2)$. $\boxdot$
	\par
	\bigskip
	\indent \textit{Proof of Theorem 1.2 :} As we mentioned above, it sufices to consider
	the case $k=\lceil (q-1)/w^2 \rceil + 1$ only. Assume $w^2>q$. Then $k=2$, thus
	$k-1 | q-1$, which makes the condition of Theorem 1.1. Now let's assume that $w|q$.
	The polynomial $f(x):=x^{q/w^2}-x$	is an additive homomorphism over $\F_q$ and
	$|\textrm{Im}f|=w^2$. By finite group theory,	there exist subgroups
	$E,F<\textrm{Im}f$ with $w$ elements, respectively, such that Im$f=E+F$.
	Further, $f \in P_k$.	Thus, from Corollary 2.1 the code RS$_k(q)$ is not
	$(w_1,w_2)$-separating. $\boxdot$ \par
	\bigskip
	From the preceding proofs, we claim that the results in
	\cite{fcsd} and \cite{mfs} can be integrated into a simpler scheme.
	We conclude this section with the following proposition that resembles
	Theorem 1.2 without proof.
	\begin{prop}
	Suppose $w_1w_2$ divides $q$. If $\RS_k(q)$ is a $(w_1,w_2)$-separating code,
	then $d>n-n/(w_1w_2)$ where $d$ is minimum distance.
	\end{prop}
	\section{New Parameter Setups}
	$w_1,w_2$ and $q$ are the parameters specifying Reed-Solomon code and its
	separation property. The aim of this section is to propose new configurations of
	them that provide Question 2 with positive answer. The underlying principle is again
	Theorem 2.1 or Corollary 2.1. \\
	\indent In this section, we suppose that $k-1 \mid q$ where
	$k=\lceil (q-1)/(w_1w_2) \rceil + 1$. Since $q$ is a prime power, there
	exists an integer $s$ such that $q/(k-1)=p^s$. One can easily check that
	$p^s$ is the largest power of $p$ which is equal or less than $w_1w_2$.
	Therefore, $s=r_1+r_2$ or $s=r_1+r_2+1$. \\
	\indent The main idea is to set $f(x):=x^{k-1}-x$, prove that Im$f$ is
	additively or multiplicatively $(w_1,w_2)$-separable, and refer to Corollary 2.1.
	It is obvious that $f \in P_k$ is an additive homomorphism over $\F_q$ and therefore
	Im$f$ is an additive group with $p^s$ elements. Thus,
	the problem is to find the setups such that Im$f$ is $(w_1,w_2)$-separable.
	The first setup is $s=r_1+r_2$.
	\begin{prop}
	If $s=r_1+r_2$, then $\textnormal{Im}f$ is additively $(w_1,w_2)$-separable.
	\end{prop}
	$Proof.$ For $|\textnormal{Im}f|=p^{r_1+r_2}$, there exist two additive subgroups $E$
	and $F$ with Im$f=E+F$ such that $|E|=p^{r_1}$ and $|F|=p^{r_2}$. Therefore,
	Im$f=(E,F)_{w_1,w_2}$. $\boxdot$ \par
	\bigskip
	\noindent	The second setup is $[w_1/p^{r_1}] \cdot [w_2/p^{r_2}] \geq p$.
	\begin{prop}
	If $[w_1/p^{r_1}] \cdot [w_2/p^{r_2}] \geq p$, \textnormal{Im}$f$ is
	additively $(w_1,w_2)$-separable.
	\end{prop}
	$Proof.$ If $s=r_1+r_2$, Im$f$ is additively $(w_1,w_2)$-separating by Proposition 3.1.
	Assume $s=r_1+r_2+1$. Then there exist three additive subgroups $E,F$ and $P$ of Im$f$ with
	Im$f=E+F+P$ such that $|E|=p^{r_1},|F|=p^{r_2}$ and $|P|=p$. Moreover, $P$ is cyclic
	since $p$ is a prime number. Let $\alpha$ be its generator. If we set
	$P_1:=\{(i \cdot [w_2/p^{r_2}])\gamma \mid 0 \leq i \leq [w_1/p^{r_1}]-1\},
	P_2:=\{j\gamma \mid 0 \leq j \leq [w_2/p^{r_2}]-1\}$ and $E'=E+P_1,F'=F+P_2$, then
	we get $P=P_1+P_2$ and Im$f=E'+F'$ since $[w_1/p^{r_1}] \cdot [w_2/p^{r_2}] \geq p$.
	Therefore, Im$f$ is additively $(w_1,w_2)$-separable. $\boxdot$ \par
	\bigskip
	\noindent The results of this section can be integrated into the following theorem.	
	\begin{thr}
	Suppose $k-1$ divides $q$ with $q/(k-1)=p^s$,
	and $s \leq r_1+r_2$ or $[w_1/p^{r_1}] \cdot [w_2/p^{r_2}] \geq p$.
	If $\RS_k(q)$ is a $(w_1,w_2)$-separating code,
	then $d>n-n/(w_1w_2)$ where $d$ is minimum distance.
	\end{thr}
	Now we state the following lemma which will be useful
	for the next section.
	\begin{lem}
	The finite field $\F_{p^s}$ is $(w_1,w_2)$-separable if at least one of the followings
	hold :
	\begin{itemize}
	\item $s \leq r_1+r_2$
	\item $[w_1/p^{r_1}] \cdot [w_2/p^{r_2}] \geq p$
	\item $w_1w_2-w_2 \geq p^s$
	\end{itemize}
	\end{lem}
	$Proof.$ We can prove in the first and second cases similarly with the propositions above
	since Im$f$ is additively isomorphism with $\F_{p^s}$. So we will only consider the third case.
	It is well known that $\F_{p^s}^{*}=\F_{p^s} \backslash \{0\}$ is a multiplicative cyclic group.
	Denote by $\alpha$ its generator. Set $E:=\{\alpha^{i(w_2-1)} \mid 0 \leq i \leq w_1-1\}$ and
	$F:=\{\alpha^j \mid 0 \leq j \leq w_2-2\}$. Then $EF=\{\alpha^i \mid 0 \leq i \leq w_1w_2-w_2\}$ and
	$\F_{p^s}^{*}=EF$ since $w_1w_2-w_2 \geq p^s$. Thus, if we set $F'=F \cup \{0\}$,
	then $\F_{p^s}=EF'$ which implies $\F_{p^s}=(E,F')_{w_1,w_2}$. $\boxdot$ \par
	\section{Application of Subspace Subcodes}
	In a linear code, there are some codewords all of whose components belong to a certain
	subset of $\F_q$. Collecting such codewords is a method of constructing a new code
	from an existing code, and it was studied in \cite{d}, \cite{j} and	\cite{hms}.
	Subfield subcode in \cite{d} is a set of codewords whose components all lie in a subfield.
	Subgroup subcodes, or subspace subcodes were introduced in \cite{j} and \cite{hms}, where
	their dimensions were estimated. Let $S$ be a $v$-dimensional subspace of $\F_q$ where
	$0 \leq v \leq m$. \textit{Subspace subcode} of Reed-Solomon code $C=\RS_k(q)$ with $S$ is
	defined to be the set of codewords from $C$ whose components all lie in $S$, and is
	denoted by $\SSRS_S(C)$.
	In this section, we will study application of subspace subcodes of Reed-Solomon
	codes to Question 2 in case $p=2$ and $q=2^m$.
	It is related to the dimensions of $\SSRS_S(C)$. \\
	\indent $\SSRS_S(C)$ is an $\F_2$-linear space. In \cite{hms}, the explicit formula
	to calculate the binary dimension of $\SSRS_S(C)$ denoted by $K(C,S)$ was proposed as
	follows :
	\begin{equation*}
	K(C,S)=\sum\limits_{j \in I_n}d_j(a_j-r_j)
	\end{equation*}
	\noindent where $I_n$ is the set consisting of the smallest integers in each modulo
	$n=2^m-1$ cyclotomic coset, $d_j$ is the cardinality of the coset containing
	$j$ denoted by $\Omega_j$, $e_j$ is the number of elements from $\Omega_j$
	lying in the set $J=\{1,2,\cdots,k\}$, $a_j=me_j/d_j$ and $r_j$'s are the
	ranks of certain $(m-v) \times a_j$ matrices called cyclotomic matrices. \par
	\indent As well as the explicit formula, they presented the following lower bound
	for the binary dimension :
	\begin{equation*}
	K(C,S) \geq L(k,v)=\sum\limits_{j \in I_n} \textrm{max} \{d_j(a_j-(m-v)),0\}
	\end{equation*}
	We will call subspace subcode
	$\SSRS_S(C)$ \textit{trivial}, provided that $K(C,S) \leq v$. Then the following
	lemma is immediately obtained.
	\begin{lem}
	Suppose that there exists a $v$-dimensional subspace of $\F_q$
	denoted	by $S$ such that the subspace subcode of $C=\RS_k(q)$ with $S$ is non-trivial.
	Then,	$C$ is not $(w_1,w_2)$-separating, provided that $S$ is $(w_1,w_2)$-separable.
	\end{lem}
	\textit{Proof.} There exists a codword $c \in \SSRS_S(C)$ with $|$Im$c| \geq 2$ because of
	non-triviality. Therefore, by Theorem 2.1, $C$ is not $(w_1,w_2)$-separating. $\boxdot$ \par
	\bigskip
	By using the lower bound $L(k,v)$, we can get the more practical result about Question 2.
	$L(k,v)$ depends on the dimension of the parent code $k$ and the dimension of the subspace
	$s$, not the subspace $S$ itself. So we can restrict to $S=\F_{2^v}$. Suppose $\SSRS_S(C)$
	is trivial, then $\SSRS_T(C)$ is also trivial where $T$ is a subspace	of $S$. Therefore
	it sufices to consider the largest power $2^v$ equal or less than $w_1w_2$.
	\begin{thr}
	Let $2^v$ be the largest power equal or less than $w_1w_2$, which satisfies
	at least one of the following conditions hold, and suppose $L(k,v)>v$.
	\begin{itemize}
	\item $v \leq r_1+r_2$
	\item $[w_1/2^{r_1}] \cdot [w_2/2^{r_2}] \geq 2$
	\item $w_1w_2-w_2 \geq 2^v$
	\end{itemize}
	If $\RS_k(q)$ is a $(w_1,w_2)$-separating code,	then $d>n-n/(w_1w_2)$ where $d$ is minimum
	distance.
	\end{thr}
	\textit{Proof.} $\SSRS_S(C)$ is non-trivial since $K(C,S) \geq L(k,v)>v$ where $S=\F_{2^v}$.
	Plus, $\F_{2^v}$ is $(w_1,w_2)$-separable by Lemma 3.1. Therefore, applying Lemma 4.1
	implies the conclusion. $\boxdot$
	\section{Examples and Conclusions}
	In this article, we presented an algebraic statement for the generalized version of
	Silverberg's open problem, and exploited it to integrate the former results.
	Besides the previous results, we could procure some new parameter setups ensuring the
	equivalence of separation and traceability properties for Reed-Solomon codes. Finally
	using the concept of subspace subcode introduced in error-correcting code theory,
	we proposed a new result when the characteristic of finite field is 2. \par
	\indent Table-1 illustrates the contributions of our work to Silverberg's open problem
	for some parameters. For each $w$ and $q$, we set $k=\lceil (q-1)/w^2 \rceil + 1$ and
	check $(w,w)$-separation property of $\RS_k(q)$ by the existing results. In each cell,
	the source of the work is written if $\RS_k(q)$ is not $(w,w)$-separating. For example, "\cite{fcsd}"
	means that non-separation is proven by Theorem 1.1, and "3.1" represents that it is
	followed by Theorem 3.1 of our paper. The symbol "*" denotes the trivial cases
	$w^2 \geq q$, and "-" stands for pending cases. \par \bigskip
	\indent \textbf{Example 1 :} Let $w=15$ and $q=256$. Then
	$k=\lceil (q-1)/w^2 \rceil + 1 = 3$. So $k-1 \mid q$. Moreover, since
	$(15/2^3)^2>2$, the condition of Theorem 3.1 holds. Therefore $\RS_k(q)$ is
	not $(w,w)$-separating. Now let $w=10$ and $q=128$. Then $(10/2^3)^2<2$.
	However, $q/(k-1)=64=2^6$, so the condition of Theorem 3.1 holds and
	$\RS_k(q)$ is	not $(w,w)$-separating. \par
	\bigskip
	\indent \textbf{Example 2 :} Let $w=12$ and $q=2048$. Then $k=16$. Theorem 3.1
	cannot be applied in this case, since $k-1=15$ divides neither $q$ nor $q-1$.
	The largest power of 2 equal or less than $w^2=144$ satisfying at least one of
	the conditions in Lemma 4.1 is $2^7=128$, for $w^2-w=132>128$.
	The modulo $n=2047$ cyclotomic coset containing 1 is
	$\Omega_1=\{1,2,4,8,16,32,64,128,256,512,1024\}$, therefore $d_j=|\Omega_1|=11$
	and $a_j=e_j=|\Omega_1 \cap J|=5$ where $J=\{1,2,\cdots,16\}$. So
	$K(C,S) \geq L(k,v) \geq \textnormal{max}\{d_1(a_1-(m-v)),0\}=11>7$, which implies
	$\RS_k(q)$ is not $(w_1,w_2)$-separating by Theorem 4.1. \par
	\bigskip
	\centerline{
	\begin{tabular}
		{|r|c|c|c|c|c|c|c|c|c|c|c|c|} \hline
		$q$ & 16 & 32 & 64 & 81 & 125 & 128 & 243 & 256 & 512 & 1024 & 2048 & 2187 \\ \hline
		$w=2$ & \cite{mfs} & \cite{mfs} & \cite{mfs} & \cite{fcsd} & \cite{fcsd} & \cite{mfs}
			& - & \cite{mfs} & \cite{mfs} & \cite{mfs} & - & \cite{mfs}\\ \hline
		$w=3$ & - & - & \cite{fcsd} & \cite{mfs} & - & -
			& \cite{mfs} & - & - & - & - & \cite{mfs}\\ \hline
		$w=4$ & \cite{fcsd} & \cite{mfs} & \cite{mfs} & \cite{fcsd} & - & \cite{mfs}
			& - & \cite{mfs} & \cite{mfs} & \cite{mfs} & \cite{mfs} & -\\ \hline
		$w=5$ & * & 3.1 & \cite{fcsd} & \cite{fcsd} & \cite{mfs} & -
			& - & - & - & - & - & -\\ \hline
		$w=7$ & * & * & 3.1 & 3.1 & - & 4.1
			& - & - & - & - & - & -\\ \hline
		$w=9$ & * & * & * & \cite{fcsd} & \cite{fcsd} & 3.1
			& \cite{mfs} & 3.1 & \cite{fcsd} & - & - & \cite{mfs}\\ \hline
		$w=10$ & * & * & * & \cite{fcsd} & \cite{fcsd} & 3.1
			& 3.1 & \cite{fcsd} & - & \cite{fcsd} & - & -\\ \hline
		$w=12$ & * & * & * & * & * & *
			& \cite{fcsd} & 3.1 & 3.1 & 3.1 & 4.1 & -\\ \hline
		$w=13$ & * & * & * & * & * & *
			& \cite{fcsd} & 3.1 & 3.1 & 4.1 & - & -\\ \hline
		$w=14-15$ & * & * & * & * & * & *
			& \cite{fcsd} & 3.1 & 4.1 & - & - & -\\ \hline
		$w=17$ & * & * & * & * & * & *
			& * & * & 3.1 & 3.1 & 3.1 & -\\ \hline
		$w=18$ & * & * & * & * & * & *
			& * & * & 3.1 & 3.1 & 4.1 & -\\ \hline
		$w=19-22$ & * & * & * & * & * & *
			& * & * & 3.1 & \cite{fcsd} & - & -\\ \hline
		$w=24$ & * & * & * & * & * & *
			& * & * & * & 3.1 & 3.1 & -\\ \hline
		$w=28-31$ & * & * & * & * & * & *
			& * & * & * & 3.1 & 4.1 & 3.1\\ \hline
		$w=34-40$ & * & * & * & * & * & *
			& * & * & * & * & 3.1 & \cite{fcsd}\\ \hline
	\end{tabular}
	} \par
	\bigskip
	\centerline{
	Table-1. Contributions to Silverberg's Problem for Some Parameters
	}	
	\bigskip	
	\noindent Thus, for a large family of Reed-Solomon codes with $2 \leq w \leq 40$ and
	$16 \leq q \leq 2187$, the separation and traceability properties are equivalent.


\begin{thebibliography}{20}
	
	\bibitem{bon}
	D. Boneh and J. Shaw, "Collusion-secure fingerprinting for digital data",
	\textit{IEEE Transactions on Information Theory}, \textbf{44}(1998), 1897-1905
	
	\bibitem{c}
	G. Cohen, "Separation and witnesses", \textit{published in International Workshop on Coding
	and Cryptography (Hunan China)}, 2009
	
	\bibitem{d}
	P. Delsarte, "On subfield subcodes of modified Reed-Solomon codes",
	\textit{IEEE Transactions on Information Theory}, \textbf{IT-21}(1975), 575-576
	 
	\bibitem{fcsd}
	M. Fernandez, J. Cotrina, M. Soriano and N. Domingo, "A note about the identifier parent
	property in Reed-Solomon codes", \textit{Computers \& Security}, \textbf{29}(2010), 628-635
	
	\bibitem{hms}
	M. Hattori, R. McEliece and G. Solomon, "Subspace subcodes of Reed-Solomon codes",
	\textit{IEEE Transactions on Information Theory}, \textbf{44}(1998), 1861-1880
	
	\bibitem{j}
	J. Jensen, "Subgroup subcodes",
	\textit{IEEE Transactions on Information Theory}, \textbf{41}(1995), 781-785
	
	\bibitem{mfs}
	J. Moreira, M. Fernandez and M. Soriano, "A note on the equivalence of the traceability
	properties of Reed-Solomon codes for certain coalition sizes", \textit{First IEEE Workshop
	on Information Forensics and Security (WIFS 2009)}, 36-40
	
	\bibitem{ssw03}
	A. Silverberg, J. Staddon and J. Walker, "Applications of list decoding to tracing traitors",
	\textit{IEEE Transactions on Information Theory}, \textbf{49}(2003), 1312-1318
	
	\bibitem{ssw01}
	J. Staddon, D. Stinson and R. Wei, "Combinatorial properties of frameproof and traceability
	codes", \textit{IEEE Transactions on Information Theory}, \textbf{47}(2001), 1042-1049
		
	\end{thebibliography}
 	\end{document}